\documentclass[english,]{IEEEtran}
\usepackage{lmodern}
\usepackage{amssymb,amsmath}
\usepackage{ifxetex,ifluatex}
\usepackage{fixltx2e} 
\ifnum 0\ifxetex 1\fi\ifluatex 1\fi=0 
  \usepackage[T1]{fontenc}
  \usepackage[utf8]{inputenc}
\else 
  \ifxetex
    \usepackage{mathspec}
  \else
    \usepackage{fontspec}
  \fi
  \defaultfontfeatures{Ligatures=TeX,Scale=MatchLowercase}
\fi
\IfFileExists{upquote.sty}{\usepackage{upquote}}{}
\IfFileExists{microtype.sty}{%
\usepackage{microtype}
\UseMicrotypeSet[protrusion]{basicmath} 
}{}
\usepackage[unicode=true]{hyperref}
\hypersetup{
            pdftitle={U2Fi: A Provisioning Scheme of IoT Devices with Universal Cryptographic Tokens},
            pdfkeywords={IoT, Universal Cryptographic Token, Provisoning},
            pdfborder={0 0 0},
            breaklinks=true}
\ifnum 0\ifxetex 1\fi\ifluatex 1\fi=0 
  \usepackage[shorthands=off,main=english]{babel}
\else
  \usepackage{polyglossia}
  \setmainlanguage[]{english}
\fi
\usepackage{graphicx,grffile}
\makeatletter
\def\maxwidth{\ifdim\Gin@nat@width>\linewidth\linewidth\else\Gin@nat@width\fi}
\def\maxheight{\ifdim\Gin@nat@height>\textheight\textheight\else\Gin@nat@height\fi}
\makeatother
\setkeys{Gin}{width=\maxwidth,height=\maxheight,keepaspectratio}
\IfFileExists{parskip.sty}{%
\usepackage{parskip}
}{
\setlength{\parindent}{0pt}
\setlength{\parskip}{6pt plus 2pt minus 1pt}
}
\providecommand{\tightlist}{%
  \setlength{\itemsep}{0pt}\setlength{\parskip}{0pt}}
\ifx\paragraph\undefined\else
\let\oldparagraph\paragraph
\renewcommand{\paragraph}[1]{\oldparagraph{#1}\mbox{}}
\fi
\ifx\subparagraph\undefined\else
\let\oldsubparagraph\subparagraph
\renewcommand{\subparagraph}[1]{\oldsubparagraph{#1}\mbox{}}
\fi

\makeatletter
\def\fps@figure{htbp}
\makeatother

\title{U2Fi: A Provisioning Scheme of IoT Devices with Universal Cryptographic
Tokens}

\author{
            \IEEEauthorblockN{Wang Kang}
        \IEEEauthorblockA{%
             \\
            Alibaba Group \\
            wangkang.wk@alibaba-inc.com}
        }

\date{}

\begin{document}
\maketitle
\begin{abstract}
Provisioning is the starting point of the whole life-cycle of IoT
devices. The traditional provisioning methods of IoT devices are facing
several issues, either about user experience or privacy harvesting.
Moreover, IoT devices are vulnerable to different levels of attacks due
to limited resources and long online duration.

In this paper, we proposed U2Fi, a novel provisioning scheme for IoT
devices. We provide a solution to make the U2F device that has been
trusted by the cloud in the distribution process, via WiFi or its side
channel, to provision the new IoT device. Further, subsequent device
settings modification, setting update, and owner transfer can also be
performed by using a U2F device that has been trusted to improve
security and provide a better user experience. This could provide
helpful user friendliness to some valuable new application scenarios in
IoT, such as smart hotel. Users could migrate the whole authentication
of smart devices into a new site by simply inserting the universal
cryptographic token into the secure gateway and authorizing by pressing
the user-presence button on the token. Besides, the relevant unbinding
process could also be done with a single cryptographic operation signed
by the cryptographic token.\\
\end{abstract}


\hypertarget{introduction}{%
\section{Introduction}\label{introduction}}

New IoT application scenarios including smart hotels, smart home,
intelligent manufacturing, brilliant factory, etc., are being developed
in the commercial market with the help of mass sales of smart speakers,
smart bulbs, smart routers, and smart sensors. However, IoT devices are
vulnerable to different levels of attacks due to limited resources and
long online duration. Moreover, the existing authentication technology
is relatively difficult to transplant to the home and office LAN
environment with a certain scale of IoT equipment due to its relatively
complicated design.

Provisioning is the first procedure in the whole life-cycle and trust
chain of IoT device, involving the exchange of sensitive credential and
plain-text passphrase of smart home infrastructure. It builds up the
cornerstone of trust in identity authentication and secure transmission
in the entire life cycle of IoT.

The traditional provisioning methods of IoT device are usually based on
temporary WiFi hotspot or WiFi side channels. Specifically, the IoT
device establishes a temporary WiFi hotspot before the smartphone
connects into this temporary network, performing related configuration.
The disadvantage is that the operation is complicated and not
user-friendly. Another commonly utilized technology that provides a
better user experience is often referred to as SmartCfg. Basically, the
underlying principle of SmartCfg is ``packet length modulation''. The
IoT device can listen to the packet length of the encrypted WiFi packet,
obtain the initial password information, and complete the access
networking. Besides, mac addresses of WiFi broadcasting packets are
being utilized as a side channel as well. Unfortunately, the credentials
of this kind of provisioning technology can be harvested by adversaries,
since the sensitive information is being transmitted over the air
without any cryptographic protection {[}1{]}. Other methods of
provisioning utilizing visual light communication between the screen and
photodiode in IoT devices are in use as well, however, this could
potentially increase the manufacturing cost.

Fortunately, there are some efforts from the market which could take
away the barrier between the cryptographic level of trust and consumer
application, such as Universal 2nd Factor token. Two-factor
Authentication (2FA) is publicly adopted as the most effective way to
reduce the incidence of online identity theft and other online fraud.
FIDO Universal 2nd Factor (U2F) {[}2{]} {[}3{]} is an open standard that
strengthens and simplifies 2FA, with native support in platforms and
browsers and has been used by Facebook {[}4{]}, Google {[}5{]}, Github
{[}6{]}, Dropbox {[}7{]}, etc. A secure element which performs
cryptographic functions and stores key pairs are embedded in each FIDO
U2F key, establishing the root of trust cryptographically. It has
several advantages over traditional password model, by introducing
challenge-response approaches to complete the verification. Furthermore,
it is USB driver-free as well as supported by multi most commonly used
browsers such as Chrome, Firefox.

In this paper, we proposed U2Fi, a novel provisioning scheme for IoT
devices. We provide a solution to make the U2F device that has been
trusted by the cloud in the distribution process, via WiFi (with IP
connection) or WiFi side channel (no IP connection). , to provision the
new IoT device. Further, subsequent device settings modification,
setting update, and owner change can also be performed by using a U2F
device that has been trusted to improve security and provide a better
user experience. This could provide helpful user friendliness to some
valuable application scenarios in IoT, such as the smart hotel. Users
could migrate the whole authentication of smart devices into a new site
by simply inserting the universal cryptographic token into the secure
gateway and authorizing by pressing the user-presence button on the
token. Besides, the relevant unbinding process could also be done with a
single cryptographic operation signed by the cryptographic token.

The rest of the paper is organized as follows. In Section II, related
background technologies are introduced, limitations of which are
discussed as well. In Section III, we introduce the design goal as well
as the architecture of U2Fi. Evaluations of a scenario in the world of
IoT are provided and discussed in Section IV.

\hypertarget{backgrounds}{%
\section{Backgrounds}\label{backgrounds}}

In this section, we will discuss several provisioning schemes that are
commonly used in the IoT world, advantages and disadvantages of which
will also be discussed. Next, we will introduce the principle and
advantages of the universal cryptographic token, and some security
events will be briefly introduced.

\hypertarget{provisioning-methods-of-iot-devices}{%
\subsection{Provisioning Methods of IoT
Devices}\label{provisioning-methods-of-iot-devices}}

Provisioning is the starting point of the whole life-cycle of IoT
Devices. It involves the exchange of sensitive credential and plain-text
passphrase of smart home infrastructure and builds up the cornerstone of
trust in identity authentication and secure transmission in the entire
life cycle of IoT.

\hypertarget{temporary-wifi-hot-spot}{%
\subsubsection{Temporary WiFi Hot Spot}\label{temporary-wifi-hot-spot}}

Most IoT devices, limited by cost and computing power, are usually
designed to be headless and lack user interaction. One of the most
common methods of provisioning is to press and hold the configuration
button on the IoT device to bring it into configuration mode.
Subsequently, the IoT device itself became a temporary WiFi hotspot.
Users use their own smartphones to access this temporary hotspot and
perform related configuration operations. This approach does not require
additional hardware and development work for IoT devices, as most IoT
devices have built-in WiFi support. For example, Google Home, Google
Chromecast, Amazon Echo, etc. all use this provisioning scheme.

However, this program has significant deficiencies. First, setting up a
temporary hotspot and switching the phone to that hotspot configuration,
and then switching back to the wireless network infrastructure used
daily, takes a long time and often leads to failure. Second, this
solution is even more powerless for the initial access scenario of
large-scale IoT devices, because its trouble is further increased.

\hypertarget{wifi-side-channel-smartcfg}{%
\subsubsection{WiFi Side Channel:
SmartCfg}\label{wifi-side-channel-smartcfg}}

Another way to provide a slightly better user experience is called
SmartCfg. SmartCfg is a provisioning scheme proposed by Texas Instrument
{[}8{]}. Its basic principle is to use some side channel features of
WiFi, such as the broadcast packet length, or the MAC address of the
broadcast packet, as the distribution method of provisioning
credentials. The advantage of this solution is that the user only needs
to press the configuration button on the IoT device to put it into
configuration mode. The mobile phone no longer needs to wait for the IoT
device to establish a temporary WiFi hotspot and then connect. However,
the user must enter the name and password of the accessed WiFi
infrastructure on the smartphone, which will then be transmitted on the
WiFi side channel in plain text or with limited encryption strength. In
recent years, some researchers have given the possibility that SmartCfg
is heavily collected in clear text transmitted during the provisioning
process {[}1{]}.

In addition, there are some restrictions in this way. For example, the
IoT device must have its WiFi network card enter a special promiscuous
mode to monitor the WiFi packets in the air. This can lead to some
compatibility issues.

\hypertarget{visible-light-communication}{%
\subsubsection{Visible Light
Communication}\label{visible-light-communication}}

There are other provisioning methods that simply introduce an additional
photodiode on the IoT device to sense the flashing intensity of the
flash on the smartphone screen, which modulates the plaintext password
required to access the WiFi. The shortcomings of this scheme are also
obvious. Due to the instability of this communication method, the
configuration fails every now and then. Furthermore, additional
photodiodes also require the introduction of new materials and
manufacturing costs, which is vital in the production process of a very
large number of IoT equipment.

\hypertarget{universal-cryptographic-tokens}{%
\subsection{Universal Cryptographic
Tokens}\label{universal-cryptographic-tokens}}

\hypertarget{fido-universal-2nd-factor}{%
\subsubsection{FIDO Universal 2nd
Factor}\label{fido-universal-2nd-factor}}

FIDO is short for Fast IDentity Online, which is an industry consortium
launched in February 2013. U2F is short for Universal 2nd Factor, which
is an open authentication standard that strengthens and simplifies
two-factor authentication using a specialized USB or NFC device. It is
initially developed by Google and Yubico.

It's now supported by Google, Facebook, Github, Dropbox, Gitlab,
Wordpress, Fastmail, Salesforce, Micro Focus, Bitbucket, dashlane,
Centrify, Duo, digidentity, RSA, PushCoin, IBM, Keeper, RCDevs,
shibboleth, Sentry, privacyIDEA, StrongAuth, Compose, AuthStack, WSO2,
gandi.net, SAASPASS and so on.

Native browser supported of Chrome was introduced in 2014 by Google.
Support for Firefox Quantum was added in November 2017. As for Safari,
there are some 3rd-party plugins.

\hypertarget{benefits-of-utilizing-u2f}{%
\subsubsection{Benefits of Utilizing
U2F}\label{benefits-of-utilizing-u2f}}

There several advantages of utilizing U2F technology, such as:

\begin{itemize}
\tightlist
\item
  Driverless: With USB-HID utilized, there's no need for an extra driver
  in the operating system. It works out of the box in most commonly used
  operating systems such as Windows, macOS, Linux, and ChromeOS.
\item
  Plugin-less: There's no extra plugin to be installed in web browsers
  such as Chrome, Firefox Quantum, Opera(based on Chromium). They are
  now providing native U2F support. Other browsers on the way.
\item
  Expandable: The underlying raw message layer can be run over Bluetooth
  Low Energy and NFC.
\item
  Privacy: It will not leak the user's identification since no biometric
  information or phone number is needed.
\end{itemize}

\hypertarget{test-of-user-presence}{%
\subsubsection{Test of User Presence}\label{test-of-user-presence}}

The Test of User Presence (TUP) is an optional function in FIDO U2F
standard to prevent inadvertent authentication when a USB U2F device is
still plugged in. Basically, it's a boolean parameter, which is signed
as well as other response data by the private key inside the USB device.

\hypertarget{secure-element}{%
\subsubsection{Secure Element}\label{secure-element}}

Typically, a well-designed U2F device should be built with a
cryptographic chip which is called `Secure Element'. A Secure Element
generally has one or more of the following abilities: a) On-chip public
and private key pair generation, and the corresponding public key can be
exported later. b) The public and private key pair are generated outside
the SE, and then the private key can be imported to the secure element.
c) Non-cryptographic storage: write-only data (One Time Programming
Zone) d) Monotonic counters. If the secure element is used as a scenario
a), the whole process can be considered as `secure-element-level' safe.
On the other hand, if scenario b) is used, during the manufacturing
process the private key could possibly be recorded.

\hypertarget{key-wrapping-mechanism}{%
\subsubsection{Key Wrapping Mechanism}\label{key-wrapping-mechanism}}

According to the FIDO U2F standard document {[}9{]}, key-wrapping
mechanism makes it possible to authenticate with unlimited numbers of
devices, allowing for inexpensive U2F devices. The Key Handle issued by
the U2F device does not have to be an index to the private key stored on
board the U2F device secure element chip. Instead, the Key Handle can
`store' (i.e., contain) the private key for the origin and the hash of
the origin encrypted with a `wrapping' key known only to the U2F device
secure element.

\hypertarget{security-evalutations-on-u2f}{%
\subsubsection{Security Evalutations on
U2F}\label{security-evalutations-on-u2f}}

Security evaluations on U2F key-wrapping {[}10{]} indicate that it could
introduce a supply-chain risk if the anti-clone counter is not well
implemented and checked. A security bulletin has been published by FIDO
alliance, and related vendors have been informed.

Another group of researchers demonstrated how to circumvent the FIDO U2F
origin check using WebUSB {[}11{]}. This kind of attack bypassed the
origin check of the web browser by setting up a phishing website
utilizing W3C WebUSB {[}12{]} to intercept U2F raw messages. It has been
fixed by the web browser vendor by blacklisting access types of WebUSB
feature.

Those attacks, as far as we know, are not related to U2F authenticators.
Thus, we should still be able to utilize the well-designed universal
authenticators as the cryptographic signature generator in IoT
scenarios.

\hypertarget{proposed-design}{%
\section{Proposed Design}\label{proposed-design}}

\begin{figure}
\centering
\includegraphics{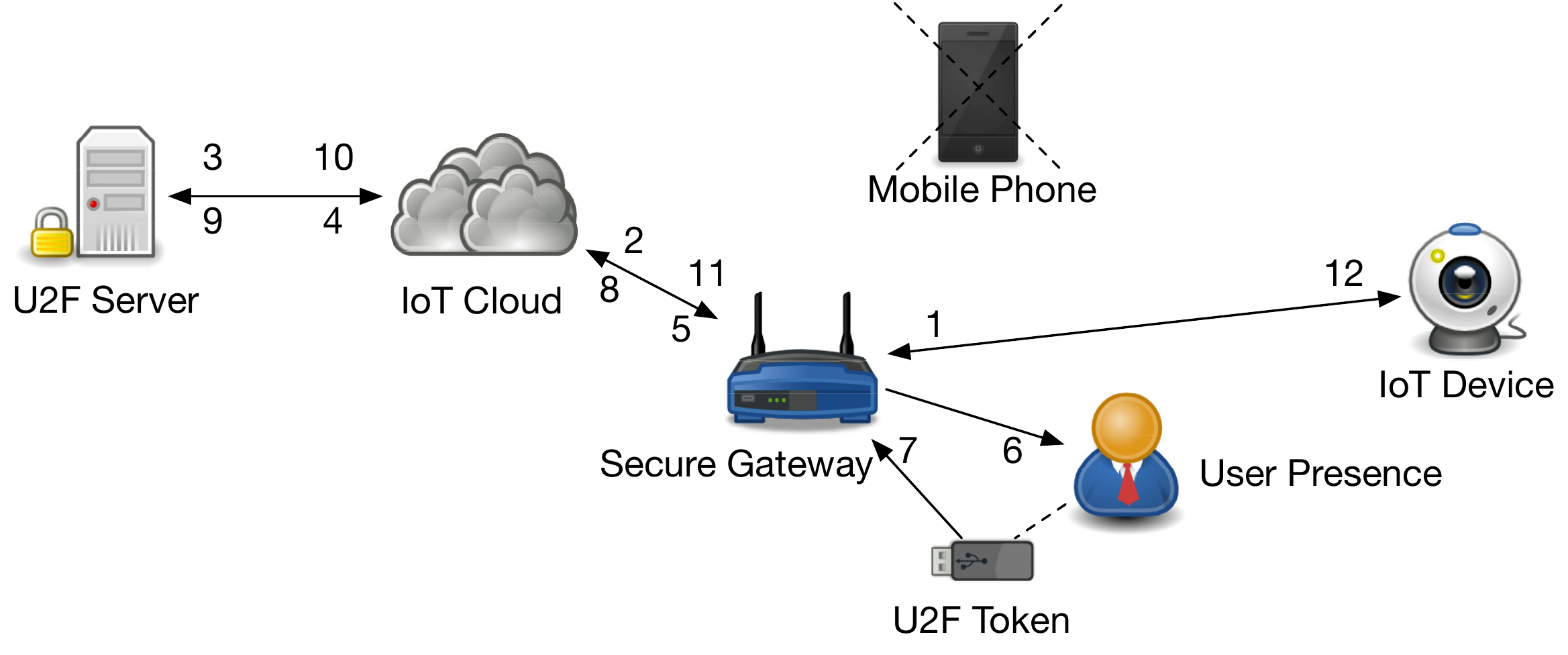}
\caption{The system architecture of proposed U2Fi provisioning scheme
for IoT devices.}
\end{figure}

\hypertarget{design-goals}{%
\subsection{Design Goals}\label{design-goals}}

Our design goal is to apply a two-factor authentication technique based
on cryptographic public key systems in a resource-constrained IoT
environment. Based on the challenge-response authentication method, the
technology provides an easy-to-develop, use, and maintain identity
authentication technology. It has the advantages of USB driver-free
support and native browser native support, which can dynamically
generate a large number of temporary keys simultaneously and meet the
multi-semantic authentication requirements.

In order to achieve this design goal, we use WiFi-based packet length
modulation technology in the U2F token access process. At present, U2F
technology only supports BTLE/NFC/USB (HID) in the physical layer, and
WiFi is not supported in the standard. The WiFi-based packet length
modulation technology enables the IoT device to obtain the initial
password information and complete the access networking by monitoring
the packet length of the encrypted WiFi packet without completing the
WiFi access.

Our design is based on U2F's IoT key nodes and authentication protocols.
IoT nodes increase the risk of being compromised and controlled by
features such as wide distribution, difficulty in operation, numerous
permissions, and long-term online. Therefore, it is necessary to further
introduce enhanced security mechanisms to improve its security
protection level. Design techniques for critical nodes and
authentication protocols rely on what users know about secrets to ensure
security, and on the other hand, use user-owned (what you have) secrets
to further enhance security. In the IoT access networking and
management, this key technology can provide an authentication mechanism
that enhances security. However, the details of its certification
agreement still need to be designed, and its security model needs to be
strictly demonstrated.

\hypertarget{architecture}{%
\subsection{Architecture}\label{architecture}}

\begin{figure}
\centering
\includegraphics{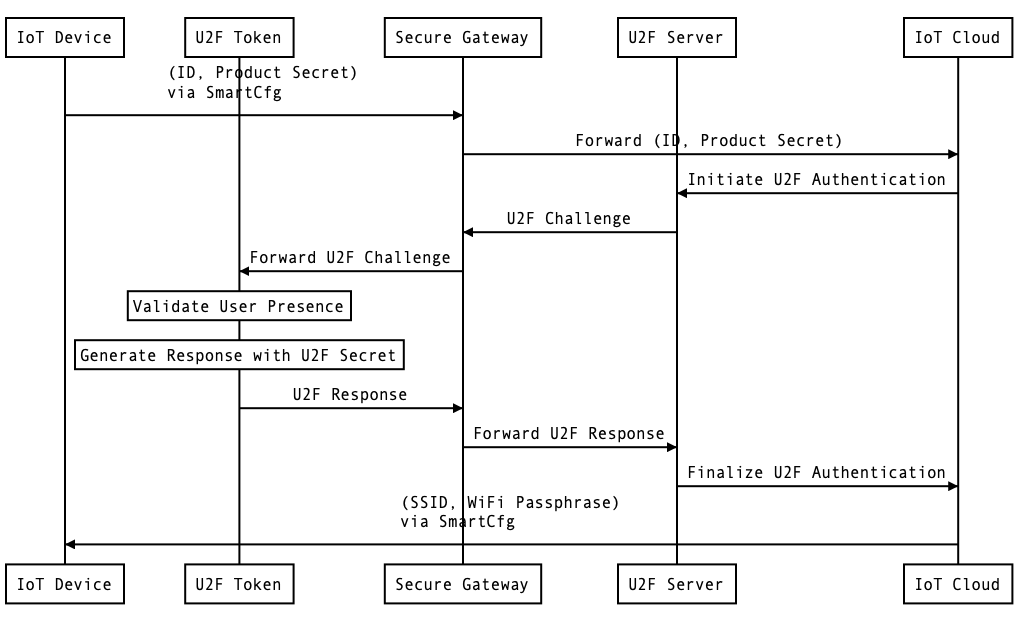}
\caption{Interaction diagram of the proposed architecture.}
\end{figure}

Typically, there are different 7 roles in our model. The IoT device,
which is a blank device without any user information. A COTS U2F key, in
possession of the user, is offline in most of the time. A mobile phone,
which is also in possession of the user, however, is online most of the
time. In our system, we will avoid any possible use of the online mobile
phone as the attestation provider. As for the cloud side, a new
authentication server providing U2F validation functionality is
introduced alongside the regular IoT service providing server.

As for the U2F key enrolling phrase, the user will initiate a normal U2F
binding process within the U2F portal website provided by the IoT
service provider. After necessary security information such as account
name, passphrase, IP address, or even biometric characteristics are
validated, the public key and key handle of the U2F token is bound to
the user's account.

As for the provisioning phrase for new registered IoT device, the user
should initiate a user identity binding request with the help of a
secure gateway. Regularly, the design could use the help of existing
user interaction method in a COTS wireless router, such as multiplexing
the WPS button. After this phrase is initiated, the IoT cloud will
contact with U2F server to generate a U2F challenge and pass it on to
the secure gateway. Then, the secure gateway will give a visual cue to
the user indicating that the U2F token should be connected within a
reasonable duration of time span. After the user has approved this
provisioning action by pressing the user-presence button on the token, a
cryptographic signature compatible with U2F response is generated and
forwarded to the relying party, i.e.~IoT cloud. Finally, the IoT cloud
will generate an access token for the IoT device after validating the
U2F response against U2F server. The entire provisioning phrase now has
been successfully proceeded.

\hypertarget{evaluations}{%
\section{Evaluations}\label{evaluations}}

In this section, we will evaluate the design of our proposed system by
implementing a new IoT scenario. We will show the necessary
implementation details. After that, some discussion of this
implementation and achievement will also be provided.

\hypertarget{scenario-description}{%
\subsection{Scenario Description}\label{scenario-description}}

This new scenario focuses on solving the problem of users in a new
environment who needs to commence the provisioning process of a bunch of
temporary IoT devices, such as smart TV, smart speaker, smart bulb,
smart power outlet, etc., and performing some high-privilege and
sensitive behaviors, such as payment. Smart routers can be implemented
as a secure gateway with reasonable small online upgrading. After
connecting these smart devices to the infrastructure, users also need to
complete the association of these devices with their own user identity.
A typical example of this scenario would be a tenant in the smart hotel.
In particular, the cryptographic token will be taken away when the user
leaves the hotel. The secure gateway can be configured to cancel the
temporary identity binding for all smart devices once the outgoing
behavior is detected.

\begin{figure}
\centering
\includegraphics{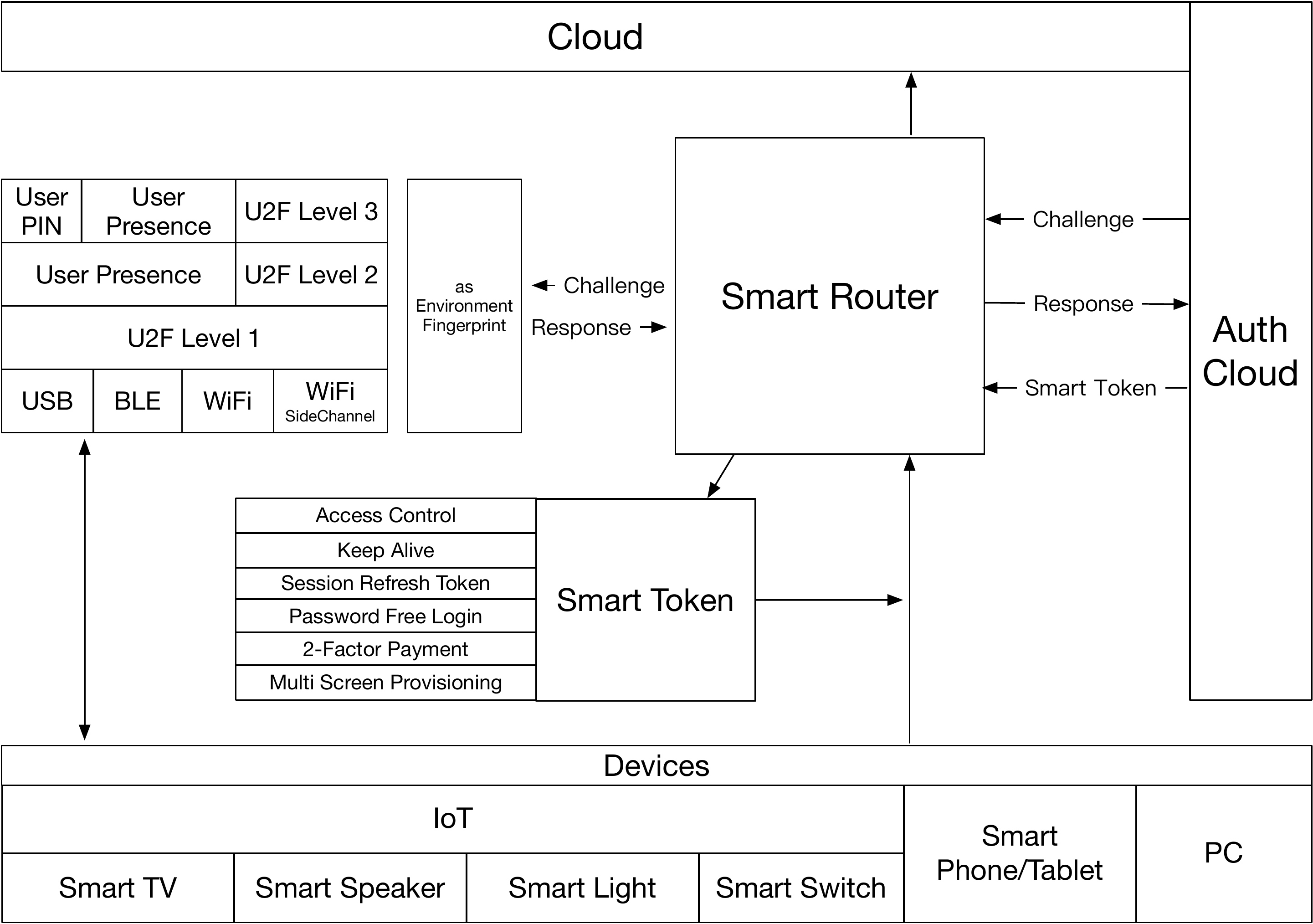}
\caption{Scenario Description}
\end{figure}

\hypertarget{implementation-of-secure-gateway}{%
\subsection{Implementation of Secure
Gateway}\label{implementation-of-secure-gateway}}

The security gateway integrates the U2Fhost forwarding module based on
the basic functions of the network connection for each IoT device, adds
U2F authentication, improves security, and implements management of
multiple IoT devices with high management efficiency.

In this part, we use the Raspberry Pi as a security gateway. The reasons
for selection are as follows: First, the Raspberry Pi supports USB
access and can carry U2F tokens; second, the Raspberry Pi comes with a
WiFi module to support various IoT devices; third, its Linux environment
can The code of the U2F-host module is integrated, and various network
protocols such as TCP/IP are supported, and the communication between
the U2F token and the U2F Server and between the IoT device and the IoT
Server can be well realized.

The U2F-host library is installed in the security gateway to exchange
information such as challenge and response with the U2F token during the
registration and authentication process. In addition, the paho-mqtt
library is installed to communicate with the cloud service platform. By
writing python code, you can exchange information between U2F tokens,
security gateways, and U2F servers.

We provide a callback function based on paho-mqtt, invoking the
\texttt{U2F-host} library. When the IoT Cloud needs to perform
registration or authentication operations, it sends a message to the
security gateway with the subject \texttt{u2f\_host}, content `register'
or `authenticate', and the security gateway discriminates the message
content and calls the corresponding function. The function of the
function is to simulate the user's registration or authentication
operation.

\hypertarget{integration-with-cloud-platform}{%
\subsection{Integration with Cloud
Platform}\label{integration-with-cloud-platform}}

The cloud service platform consists of two parts: U2F Server and IoT
Cloud. It can be implemented on a PC by writing the corresponding python
code. The U2F Server is a server that implements U2F registration and
authentication. Support U2F registration/authentication operations. When
a U2F Zero token is registered, the U2F Server will save the public key.
When authentication is required, the U2F Server uses the public key to
decrypt the encrypted data sent by the U2F host to complete the
authentication. After registration/authentication, the U2F Server sends
a message to the IoT Cloud indicating whether the
registration/authentication was successful. After IoT Cloud receives
this message, it can proceed. We provide a callback library for the U2F
Server to serve those two functions.

The IoT Cloud is responsible for recording the registration information
of the IoT device and allowing the user to remotely monitor the running
status of the IoT device. User operations on the IoT Cloud can be
divided into three categories: U2F registration operations, important
operations (requires U2F authentication), and non-critical operations
(no U2F authentication required).

\begin{itemize}
\tightlist
\item
  When the user initiates a U2F registration operation, the IoT Cloud
  will send an instruction to the security gateway to instruct it to
  start registration, and when the authentication is completed, ask the
  user which operations are set as important operations.
\item
  When the user initiates an important operation, the IoT Cloud first
  sends an instruction to enable the U2F authentication to the security
  gateway. After the authentication is completed, the IoT Cloud sends
  the operation command to the IoT device.
\item
  When the user initiates a non-critical operation, the IoT Cloud sends
  the operation command directly to the IoT device.
\end{itemize}

In the demonstration system, IoT Cloud can control the IoT device and
realize the function of turning on the camera, which requires U2F
authentication. The camera is initialized after \texttt{register} and
\texttt{open} instruction input by the user.

\hypertarget{emulation-of-iot-devices}{%
\subsection{Emulation of IoT Devices}\label{emulation-of-iot-devices}}

In the demo system, the IoT device, emulated by a Raspberry Pi, will
open the camera after receiving the command. The \texttt{picamera}
library is utilized to control the onboard camera, and the
\texttt{paho-mqtt} library is used to communicate with the IoT cloud.
The core code is a callback function based on \texttt{paho-mqtt}, as
shown in the following pseudo-code. When the message
\texttt{U2F\_AUTHENTICATED} sent by the IoT cloud is received, the
camera is turned on and the preview mode is entered.

\hypertarget{discussion}{%
\subsection{Discussion}\label{discussion}}

\begin{figure}
\centering
\includegraphics{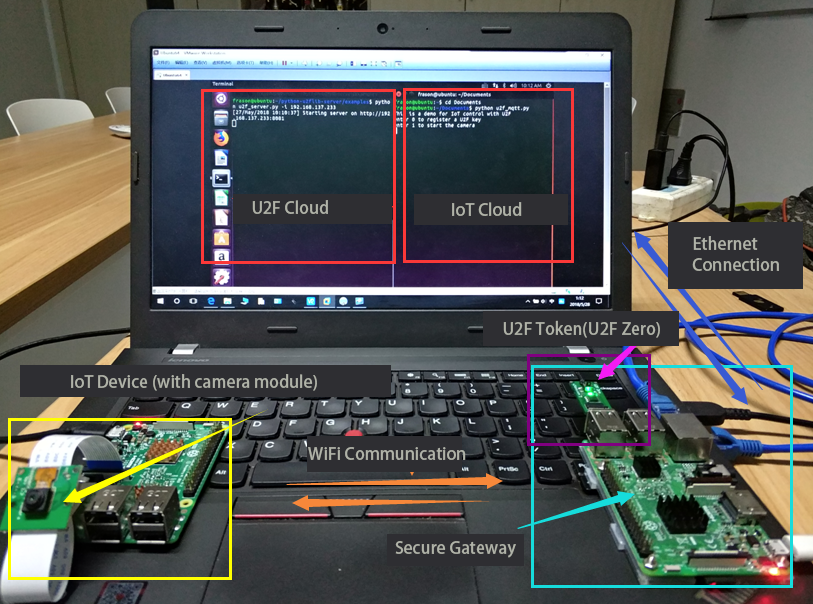}
\caption{Demo system}
\end{figure}

\begin{figure}
\centering
\includegraphics{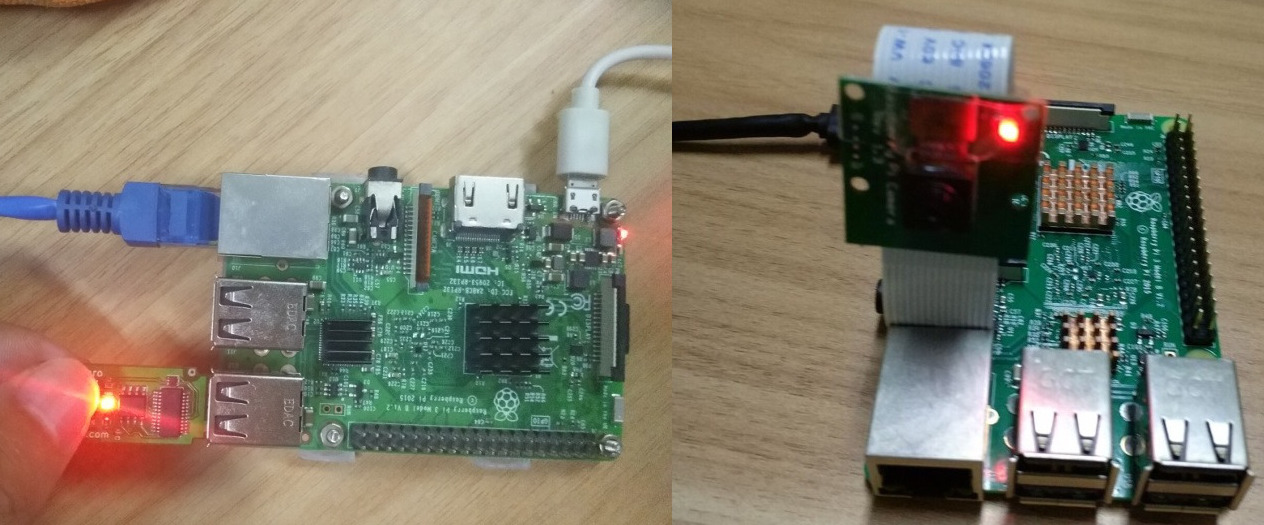}
\caption{Validation of U2F user presence. \textbar{} IoT Camera
registered accordingly.}
\end{figure}

It can be seen from the above evaluations that the system can
successfully add U2F authentication to the communication between the IoT
device and the cloud server, and realize the verification of the
physical authorization of the U2F token to the user entity.

In the existing IoT system, the IoT Cloud does not need to be
authenticated for the IoT device. If the attacker steals the user's ID
and password, the IoT Cloud can be logged in to and send malicious
instructions to the IoT device. In this system, before the cloud service
sends important instructions to the device, it needs to wait for the
user to perform U2F authentication operation (ie, press the button of
the U2F token inserted on the security gateway). After the U2F
authentication succeeds, the relevant commands can pass the security.
The gateway sends the IoT device to ensure the security of communication
between the IoT device and the cloud server.

Besides, we provide an improved SmartCfg distribution method. In the
traditional distribution network solution and the mainstream SmartCfg
distribution solution, the mobile device App and the IoT device need to
be connected to each other, and then the SSID and password of the router
are transmitted through the WiFi packet length modulation. In this work,
the IoT device's network connection and U2F authentication are combined,
and the U2F token on the security gateway can be operated securely,
which greatly simplifies the distribution process.

At the same time, the combination of U2F standard and IoT security
certification. U2F, as an emerging network security authentication
method, has been widely used in many large-scale services, but
basically, it is aimed at network security. Combined with the current
hot concept of IoT security authentication, we have successfully
combined the two to achieve two-factor security authentication on the
IoT platform, thereby using the security gateway (key node) to
authenticate multiple IoT devices mounted. It achieves the purpose of
secure communication and enhances the user's operating experience, which
has great application prospects.

This could provide helpful user friendliness to some valuable new
application scenarios in IoT, such as the smart hotel. Users could
migrate the whole authentication of smart devices into a new site by
simply inserting the universal cryptographic token into the secure
gateway and authorizing by pressing the user-presence button on the
token. Besides, the relevant unbinding process could also be done with a
single cryptographic operation signed by the cryptographic token.

\hypertarget{conclusion}{%
\section{Conclusion}\label{conclusion}}

Our proposed solution includes a new SmartCfg encoding method to combine
U2F with IoT security certification. Further, a feasible solution is
proposed in the field of IoT security certification and a demonstration
system is given. Considering smart hotel, smart home and smart office
scenarios, we can apply this solution to IoT devices and implemented the
security management of batch devices, and improve the user's operating
experience while ensuring security. The solution requires multi-party
support for IoT terminals, security gateways (routers), U2F physical
tokens, and commercial cloud services, which has great application
prospects and a broad market.

\hypertarget{references}{%
\section*{References}\label{references}}
\addcontentsline{toc}{section}{References}

\hypertarget{refs}{}
\leavevmode\hypertarget{ref-li_passwords_2018}{}%
{[}1{]} C. Li, Q. Cai, J. Li, H. Liu, Y. Zhang, D. Gu, and Y. Yu,
``Passwords in the Air: Harvesting Wi-Fi Credentials from SmartCfg
Provisioning,'' in \emph{Proceedings of the 11th ACM Conference on
Security \& Privacy in Wireless and Mobile Networks - WiSec '18}, 2018,
pp. 1--11.

\leavevmode\hypertarget{ref-srinivas2015universal}{}%
{[}2{]} S. Srinivas, D. Balfanz, E. Tiffany, A. Czeskis, and F.
Alliance, ``Universal 2nd factor (u2f) overview,'' \emph{FIDO Alliance
Proposed Standard}, pp. 1--5, 2015.

\leavevmode\hypertarget{ref-balfanz2015fido}{}%
{[}3{]} D. Balfanz, ``FIDO u2f raw message formats,'' \emph{FIDO
Alliance Proposed Standard}, pp. 1--10, 2015.

\leavevmode\hypertarget{ref-gandheleexpanding}{}%
{[}4{]} S. Gandhele, ``Expanding universal second factor (u2f) to
non-browser applications.''

\leavevmode\hypertarget{ref-grossklags_security_2017}{}%
{[}5{]} J. Lang, A. Czeskis, D. Balfanz, M. Schilder, and S. Srinivas,
``Security Keys: Practical Cryptographic Second Factors for the Modern
Web,'' in \emph{Financial Cryptography and Data Security}, vol. 9603, J.
Grossklags and B. Preneel, Eds. Berlin, Heidelberg: Springer Berlin
Heidelberg, 2017, pp. 422--440.

\leavevmode\hypertarget{ref-dautermanmaking}{}%
{[}6{]} E. Dauterman, H. Corrigan-Gibbs, D. Mazières, D. Boneh, and D.
Rizzo, ``Making u2f resistant to implementation bugs and supply-chain
tampering.''

\leavevmode\hypertarget{ref-lang2016security}{}%
{[}7{]} J. Lang, A. Czeskis, D. Balfanz, M. Schilder, and S. Srinivas,
``Security keys: Practical cryptographic second factors for the modern
web,'' in \emph{International conference on financial cryptography and
data security}, 2016, pp. 422--440.

\leavevmode\hypertarget{ref-noauthor_cc3100_nodate}{}%
{[}8{]} ``CC3100 Provisioning Smart Config - Texas Instruments Wiki.''
\url{http://processors.wiki.ti.com/index.php/CC3100_Provisioning_Smart_Config}.

\leavevmode\hypertarget{ref-noauthor_universal_nodate}{}%
{[}9{]} ``Universal 2nd Factor (U2F) Overview.''
\url{https://fidoalliance.org/specs/fido-u2f-v1.0-nfc-bt-amendment-20150514/fido-u2f-overview.html\#allowing-for-inexpensive-u2f-devices}.

\leavevmode\hypertarget{ref-wang_u2fishing:_nodate}{}%
{[}10{]} K. Wang, ``U2Fishing: Potential Security Threat Introduced by
U2F Key-Wrapping Mechanism,'' p. 8.

\leavevmode\hypertarget{ref-vervier_oh_nodate}{}%
{[}11{]} M. Vervier and M. Orrù, ``Oh No, Where's FIDO? - A Journey into
Novel Web-Technology and U2F Exploitation.''.

\leavevmode\hypertarget{ref-noauthor_webusb_nodate}{}%
{[}12{]} ``WebUSB API.'' \url{https://wicg.github.io/webusb/}.

\end{document}